\documentclass[12pt]{iopart}

\usepackage{iopams}
\usepackage{epsf}
\def\Journal#1#2#3#4{{#1} {\bf #2}, #3 (#4)}

\def\JDG{\em J. Diff. Geom.}
\def\CQG{\em Class. Quantum Grav.}
\def\JPA{\em J. Phys. A: Math. Gen.}
\def\PRD{\em Phys. Rev. D }
\def\GRG{\em Gen. Rel. Grav.}

\def\JMP{\em J. Math. Phys.}

\def\CMP{\em Commun. Math. Phys.}

\def\PRL{\em Phys. Rev. Lett.}

\def\ANYAS{\em Ann. N. Y. Acad. Sci.}

\def\NC{\em Nuovo Cimento}
\def\AM{\em Ann. Math.}

\newcommand{\bm}[1]{\mbox{\boldmath $#1$}}

\def\espaitemps{({\cal V},g)}
\def\varietat{{\cal V}}

\def\xiv{\vec \xi }

\def\lie{{\pounds}}
\def\xif{\mbox{\boldmath $ \xi $}}
\def\etaf{\mbox{\boldmath $ \eta $}}

\def\t{\tau}

\def\f{\varphi}

\def\S{\Sigma}
\def\r{{\rm I\!R}}
\def\di{{\rm div}}

\def\be{\begin{equation}}
\def\ee{\end{equation}}
\def\bea{\begin{eqnarray}}
\def\eea{\end{eqnarray}}
\def\bean{\begin{eqnarray*}}
\def\eean{\end{eqnarray*}}

\def\proof{\noindent{\em Proof.\/}\hspace{3mm}}
\def\fin{\hfill \rule{2.5mm}{2.5mm}\\ \vspace{0mm}}

\newtheorem{prop}{Proposition}
\newtheorem{theorem}{Theorem}

\newtheorem{lemma}{Lemma}
\newtheorem{coro}{Corollary}[theorem]

\begin{document}
\title{Trapped surfaces and symmetries}
\author{Marc Mars$^1$ and 
Jos\'e M.M. Senovilla$^2$}
\address{$^1$ \'{A}rea de F\'{\i}sica Te\'orica, Facultad de Ciencias,
Universidad de Salamanca, Plaza de la Merced s/n, 37008 Salamanca, Spain\\ 
$^2$ Departamento de F\'{\i}sica Te\'orica, Universidad del Pa\'{\i}s Vasco, 
Apartado 644, 48080 Bilbao, Spain} 
\eads{\mailto{marc@usal.es} and \mailto{wtpmasej@lg.ehu.es}}
\begin{abstract} 
We prove that strictly stationary spacetimes cannot contain closed trapped 
nor marginally trapped surfaces. 
The result is purely geometric and holds in arbitrary dimension. Other 
results concerning the interplay between (generalized) symmetries and trapped 
submanifolds are also presented.
\end{abstract} 



\vspace{1cm}
The importance of Penrose's concept of closed trapped surface, its 
usefulness, and the versatility of their applications is indubitable. 
To mention just a few outstanding situations where it has 
been essential, we can cite the development of the singularity 
theorems (see e.g. \cite{P2,HP,HE,S1}),
the general analysis of gravitational collapse and formation
of trapped surfaces and black holes \cite{GC}, the study of the cosmic 
censorship hypothesis \cite{CC} and the related Penrose inequality 
\cite{PI}, or 
the numerical analysis of the Cauchy development of apparently innocuous 
initial data (e.g. \cite{Num} and references therein.)

In General Relativity, a {\em trapped surface} is a two-dimensional 
imbedded spatial surface such that the product of the traces 
of their two future-directed null second fundamental forms is 
everywhere positive. A more 
physical way of saying the same is that the two families of 
future-directed null geodesics orthogonal to the surface are, 
at least initially, simultaneously converging (or diverging). 
This concept can be easily translated 
to submanifolds of co-dimension 2 in any Lorentzian manifold $\espaitemps$
of arbitrary dimension $D$ (see e.g. \cite{Kr,S2}). 

A spacetime will be called `strictly stationary' if it contains
a Killing vector field which is timelike everywhere. The term `closed' 
is used for {\em compact without boundary}.
In this letter we want to present a simple direct proof of the following 
result: closed trapped surfaces are absent in strictly stationary spacetimes. 
Obviously the same holds in those regions of a spacetime where there
is a timelike Killing vector. A number of related results, some of them of more 
generality ---concerning 
trapped submanifolds of any co-dimension or more general symmetries---, 
some others more specific ---concerning static spacetimes---, 
and the formulae leading to them, are 
also presented. Our main result might seem intuitively obvious at first, 
as the area element of surfaces does not change along the timelike direction 
defined by 
the Killing vector in a strictly stationary spacetime. However, it is not 
straightforward: {\em there are} trapped surfaces in 
strictly stationary, and in static, spacetimes, even in the simplest examples. 
For instance, flat spacetime has trapped surfaces: for fully
explicit examples see Example 4.1 in \cite{S1}, p. 776. Thus, the 
requirement that the surface be closed is indispensable for our result.

To our purposes, and in order to keep the full generality, the best 
adapted definition of trapped surfaces is the one characterizing them 
through the use of their mean curvature vector $\vec H$, \cite{Kr,S2}. 
The virtue of this characterization is that it can be easily generalized
to imbedded submanifolds of any co-dimension. Thus, let $S$ be any 
$d$-dimensional $C^2$ imbedded spacelike submanifold in a causally orientable 
spacetime $\espaitemps$ (metric $g$ with signature (-,+,\dots,+))
and let $\vec{H}$ be its mean curvature vector
\cite{Kr,Jost,O}\footnote{The mean curvature vector can be defined
according to several conventions. Our definition is such that
$\vec{H}$ is the trace of the second fundamental form vector, without
dividing by the dimension of the submanifold, and that it points outwards
for a sphere in Euclidean space.}. We will assume, 
without further explicit mention, that all submanifolds $S$ are 
orientable. As usual we take $\vec{0}$ to be
null (future and past) but not timelike or spacelike.
In this letter we shall say that $S$ is
\begin{enumerate}
\item {\em future trapped} if $\vec H$ is timelike and 
future-pointing all over $S$. Similarly for {\em past trapped}.
\item {\em nearly future trapped} if $\vec H$ is causal and 
future-pointing all over $S$ and timelike at least at a point of $S$,
and correspondingly for {\em nearly past trapped}.
\item {\em marginally future trapped} if $\vec H$ is null and
future-pointing 
all over $S$ and non-zero at least at a point of $S$, 
and analogously for {\em marginally past trapped}.
\item {\em extremal} or {\em symmetric} if $\vec H =\vec 0$ all over $S$.
\item {\em absolutely non-trapped} if $\vec H$ is spacelike all over $S$.
\end{enumerate}
{\bf Remarks} 
\begin{itemize}
\item Definitions (i), (iv), (v) coincide with the standard ones in
co-dimension two, while in that case (iii) is more general than the standard 
concept (e.g. \cite{HE,S1}) because both expansions may
be non-identically zero on $S$ but with a vanishing product everywhere. 
Nevertheless, all the standard, non-extremal, marginally trapped 
$(D-2)$-surfaces are included in (iii). 
\item The above nomenclature is somewhat peculiar for the cases of co-dimension
one or dimension one. In the 
first possibility $S$ is a spacelike hypersurface and, concentrating
on the future case for concreteness,
(i)  corresponds to a positive expansion everywhere on $S$, 
(ii) to a  non-negative expansion which is positive at some point,
(iv) to a maximal $S$, 
while (iii) and (v) are impossible. Similarly, $S$ represents a spacelike
curve when $d=1$ and
the concept of future trapping means here that the 
proper acceleration vector is past timelike  along the curve for (i), 
null, past-pointing and not identically zero for (iii),
and that $S$ is a geodesic for (iv). 
\end{itemize}

The main tool in our reasoning is the known formula of the variation of the 
volume of any submanifold \cite{Jost}. We present here a simple 
brief derivation adapted to our goals. Let $\S$ be any smooth 
abstract $d$-dimensional manifold and let $\Phi :\S \longrightarrow 
\varietat$ be a $C^2$ imbedding into the spacetime. We shall speak 
indistinctly of $\S$ and its image in the spacetime
$\Phi(\S)\equiv S$ if no confusion arises. By means of the 
push-forward $\Phi'$ one can define a set of $d$ linearly independent
vector fields $\vec{e}_{A}$ on $S$ ($A,B,\ldots =D-d,\ldots, D-1$) which are 
tangent to $S$. Let $\gamma_{AB}=g(\vec{e}_{A},\vec{e}_{B})$ and 
$\etaf_{S}$ denote the first fundamental form and the corresponding canonical 
volume element $d$-form of $S$ in $\espaitemps$, respectively. Thus, the 
$d$-volume of $S$ is simply
$$
V_{S}\equiv \int_{S} \etaf_{S} \, .
$$
We wish to know the variation of $\etaf_{S}$, and of $V_{S}$, when we perform 
a deformation of the submanifold $S$. To that end, let $\xiv$ be an
arbitrary $C^1$ vector field on $\varietat$ defined on a neighbourhood of $S$.
$\xiv$  generates a local one-parameter group $\{\f_{\t}\}_{\t\in I}$ of local
transformations, where $\t$ is the canonical parameter and 
$I\subset\r$ is an interval of the real line containing $\t =0$. We 
define a one-parameter family of surfaces $S_{\t}\equiv \f_{\t}(S)$ 
in $\varietat$, with corresponding imbeddings 
$\Phi_{\t}:\S\rightarrow \varietat$ given by $\Phi_{\t}=\f_{\t}\circ 
\Phi$. Observe that $S_{0}=S$. The corresponding first fundamental 
forms are simply 
$\gamma_{AB}(\t)=(\f^{*}_{\t}g)(\vec{e}_{A},\vec{e}_{B})$, with 
associated canonical volume element $d$-forms $\etaf_{S_{\t}}$
(a $*$ denotes the pull-back.)
Then, it is a matter of simple calculation to get
\be
\left.\frac{d\etaf_{S_{\t}}}{d\t}\right|_{\t=0}=
\frac{1}{2}\mbox{tr}_{S}\left[\Phi^{*}(\lie_{\xiv} \, g)\right]
\etaf_{S}=
\frac{1}{2}\gamma^{AB}(\lie_{\xiv} \, g)(\vec{e}_{A},\vec{e}_{B})\, 
\etaf_{S},
\label{dereta}
\ee
where $\mbox{tr}_{S}$ denotes the trace in $S$ with respect to the induced
metric and $\lie_{\xiv}$ is 
the Lie derivative with respect to $\xiv$. Another straightforward 
computation using the standard formulae relating the connections on
$\espaitemps$ and on $(S,\gamma)$ (see, e.g. \cite{O}) leads to
\be
\frac{1}{2}\mbox{tr}_{S}\left[\Phi^{*}(\lie_{\xiv} \, g)\right]=
\di \vec{\bar{\xi}} + (\xiv\cdot \vec{H})|_{S} \label{vareta}
\ee
where $(\, \cdot \,)$ is the $g$-scalar product, 
$\di$ is the divergence operator on $S$ and $\vec{\bar{\xi}}$ is 
the projection of $\xiv$ to $S$, that is to say, 
$\bar{\xif}=\Phi^{*}\xif$ so that 
$\bar{\xi}_{A}=(\xiv\cdot\vec{e}_{A})|_{S}$. These are the formulas we 
need, but for completeness we remark that from (\ref{dereta}) and 
(\ref{vareta}) one easily derives the expression for the variation of 
$d$-volume:
$$
\left.\frac{dV_{S_{\t}}}{d\t}\right|_{\t=0}=\int_{S}\left(\di \vec{\bar{\xi}} + 
(\xiv\cdot\vec{H})\right)\, \etaf_{S} \,\, .
$$
\begin{lemma}
If $\xiv$ is a Killing vector and $S$ is a closed imbedded
submanifold, then
$$
\int_{S} (\xiv\cdot\vec{H}) \, \etaf_{S} =0 \, .
$$
\label{res1}
\end{lemma}
\proof If $\lie_{\xiv} \, g=0$, from (\ref{vareta}) we obtain 
$(\xiv\cdot\vec{H})=-\di \vec{\bar{\xi}}$. Integrating this function 
over $S$, and using Gauss' theorem, the result follows as $S$ is 
compact without boundary.\fin
{\bf Remark} The geometric reason behing this result is that the $d$-volume
of an imbedded
closed submanifold is invariant under isometries of the spacetime.
Note, however, that the proof above also holds if just the part of 
$\lie_{\xiv} \, g$ tangent to $S$ is traceless. 

We arrive at our main result.
\begin{theorem}
If $\espaitemps$ is strictly stationary, then there are no 
marginally trapped, nearly trapped, nor trapped closed imbedded
submanifolds in the spacetime. 
\label{fund}
\end{theorem}
\proof We can assume that the timelike Killing 
vector, say $\xiv$, is future pointing. Then, if $\vec H$ pointed to the future 
(past) everywhere on $S$, $(\xiv\cdot\vec{H})$ would be non-positive 
(non-negative) all over $S$, in contradiction with Lemma \ref{res1} 
unless $\vec H=\vec 0$ everywhere on $S$.\fin
{\bf Remarks} 
\begin{itemize} 
\item Observe that this theorem implies, in particular, that 
there cannot be closed spacelike curves with a causal future acceleration 
vector in strictly stationary spacetimes, except for the extremal case of closed
spacelike geodesics. Similarly, any closed spacelike hypersurface in a 
strictly stationary spacetime must have an expansion which changes sign, 
unless the hypersurface is maximal. The latter case is certainly possible.
In fact, asymptotically flat maximal slices always exist in
strictly stationary,
asymptotically flat spacetimes \cite{C1} (see also \cite{C2} for 
generalizations to asymptotic stationarity).

\item The extreme possibility $\vec H=\vec 0$ can in fact happen in any dimension
or co-dimension. Apart from what was said in the previous remark
regarding closed geodesics 
or maximal hypersurfaces, there are also simple examples
for the relevant case of co-dimension two. For instance, choose any 
$(D-1)$-dimensional proper Riemannian manifold containing a minimal 
surface $S$ and take its direct product with $(\r,-dt^2)$. A
more  sophisticated yet explicit example 
for non-static spacetimes was presented by Newman in \cite{N} 
for the famous G\"odel spacetime (see also \cite{Kr}, pp.393-405).
\end{itemize}
By similar reasonings one can prove

\begin{theorem}
\label{sec}
Let $S$ be closed and $\xiv$ any Killing vector which is null 
and nowhere vanishing
on $S$. Then, $S$ cannot be trapped nor nearly trapped, and can be
marginally trapped
if and only if its mean curvature vector points along the same 
direction as $\vec{\xi}$ everywhere.
\end{theorem}

\proof Since $\vec{\xi}$ does not vanish on $S$
it follows that it is either
future null or past null everywhere on $S$. For a (marginally, nearly)
trapped $S$, $(\vec{\xi} \cdot \vec{H})$ is either non-negative or 
non-positive, therefore $\int_{S} (\xiv\cdot\vec{H}) \, \etaf_{S} =0$
implies $(\xiv \cdot \vec{H}) |_S=0$, so that $\vec H\propto \xiv$. \fin

{\bf Remark} A consequence of this theorem is that spacetimes representing 
pp-waves (which admit
a covariantly constant and nowhere zero null Killing vector field
$\vec{\xi}\,$, see e.g. \cite{Exact})
do not admit closed
spacelike hypersurfaces which are everywhere expanding (or contracting). 
In addition, there cannot be closed trapped or nearly trapped
submanifolds and
any closed marginally trapped submanifold $S$ must necessarily have a
mean curvature vector parallel to $\vec{\xi}$.
Therefore, $S$ must be contained 
in one of the null hypersurfaces orthogonal (and tangent) to $\vec{\xi}$.

Combining the previous theorems we get:
\begin{coro}
\label{cor1}
Let $\xiv$ be any Killing vector on $\espaitemps$ and $S$ be any 
marginally trapped, nearly trapped, or trapped, closed submanifold.
Then, (i) none of 
the connected components of $S$ can be fully contained in the 
region where $\xiv$ is timelike; (ii) $S$ can be within the subset where
$\xiv$ is null and non-zero only in the case that $S$ is 
marginally trapped (and $\xif \wedge \bm{H} =0$.)
\end{coro}

{\bf Remark} It can be easily proven that Theorems \ref{fund} and \ref{sec}
and Corollary \ref{cor1}
also hold for immersed submanifolds. The details,
however, are notationally cumbersome and we prefer to omit them.

The remark after lemma \ref{res1} allows for a relaxation of the 
Killing property for $\xiv$. As an illustrative example, among many 
others, we can use 
the so-called Kerr-Schild vector fields introduced in \cite{CHS}. 
These are vector fields satisfying 
$$\lie_{\xiv} \, g\propto \bm{k}\otimes \bm{k}$$ 
for a null vector field $\vec k$. Similarly, we could also use for $\xiv$ the 
more elaborated ``causal symmetries'' recently introduced in \cite{GS}. 
For the Kerr-Schild case we have 
$\frac{1}{2}\gamma^{AB}(\lie_{\xiv} \, g)(\vec{e}_{A},\vec{e}_{B})\propto 
\gamma^{AB}(\vec k \cdot \vec{e}_{A})(\vec k \cdot \vec{e}_{B})$ which has the 
sign of the proportionality function, or vanishes if $\vec k$ is 
orthogonal to $S$. In the former case, and if the proportionality 
factor has a sign on $S$, one can derive properties about the future or
past trapping of the sets in a manner analogous to the one we shall use 
later for conformal motions. In the latter case, theorems \ref{fund} and
\ref{sec} hold just the same for these Kerr-Schild vector fields. 
An explicit example of this situation is provided by the Vaidya 
radiating spacetime which has a proper Kerr-Schild motion $\xiv$ (see 
\cite{CHS} p.667). This $\xiv$ is partly timelike, partly spacelike, and 
null at the frontier, which is the apparent horizon. One can easily 
check in this simple spacetime that the
spherically symmetric closed
trapped surfaces lie entirely in the spacelike region of $\xiv$, and 
that the apparent horizon is formed by marginally trapped spheres 
(see e.g. Example 4.2 in \cite{S1}.) Actually, the Kruskal extension 
of Schwarzschild spacetime is contained here as the subcase with 
constant mass function, and then the vector field $\xiv$ coincides with 
the asymptotically stationary Killing vector.

Even if the part of $\lie_{\xiv} \, g$ tangent to $S$ is not trace-free 
one can obtain information from formula (\ref{vareta}). Take the 
relevant example of conformal Killing vectors 
$$
\lie_{\xiv} \, g=2\phi g 
$$
so that 
$\frac{1}{2}\gamma^{AB}(\lie_{\xiv} \, g)(\vec{e}_{A},\vec{e}_{B})=\phi d$. 
When $S$ is closed, integration of (\ref{vareta}) provides
$$
\int_{S}\phi \, \etaf_{S} =\frac{1}{d} \int_{S} (\xiv\cdot\vec{H}) \,
\etaf_{S} .
$$
If $\xiv$ is timelike and $\phi|_{S}$ has a sign ---this 
includes, in particular, all timelike homothetic vectors---, then
any (marginally, nearly) trapped closed $S$ must have a fixed causal 
orientation for the trapping. If $\phi|_{S}\geq 0$, say, then $\vec H$ 
must be oppositely oriented to $\xiv$. Furthermore, extremal 
submanifolds are forbidden. As an illustrative example, 
take the obvious future timelike conformal Killing vector field of 
Robertson-Walker 
cosmological spacetimes. Application of the above reasoning 
immediately leads to the conclusion that, for {\em expanding} Robertson-Walker
cosmologies, all possible closed trapped surfaces must be {\em 
past-trapped}, and future trapped if the spacetime is contracting, and 
in neither case there can be minimal closed surfaces, nor maximal 
closed hypersurfaces. Of 
course, this corroborates the usual elementary knowledge (see e.g. 
\cite{S1} p.779.)

Let us finally make some considerations about the particular case 
of strictly
{\em static} spacetimes. In this case, there is a timelike Killing 
vector $\xiv$ which is integrable, that is $\xif\wedge d\xif =0$ so 
that $\xiv$ is orthogonal to a family of spacelike hypersurfaces. 
Being $\xiv$ a Killing vector without rotation, these hypersurfaces 
are all {\em time symmetric} in the sense that their second fundamental forms 
in $\espaitemps$ vanish. A partial generalization of this result to sets of 
arbitrary co-dimension is the following.
\begin{prop}
\label{prop1}
Let $\xiv$ be a hypersurface-orthogonal timelike Killing vector. Then, 
all submanifolds contained in any one of its orthogonal hypersurfaces have 
a mean curvature vector which is everywhere spacelike or zero.
\end{prop}
\proof Let $S$ be the submanifold. As $\xiv$ is orthogonal to $S$, 
the tangent projection $\vec{\bar{\xi}}$ appearing in (\ref{vareta}) 
vanishes. As the lefthand side of that equation also vanishes we 
get $(\xiv\cdot\vec H) |_S=0$, so that $\vec H$ must be 
spacelike or zero everywhere.\fin

This result also follows directly by using the time-inversion isometries
of a static spacetime. However, the proof we have presented (and therefore the 
proposition)
holds for arbitrary submanifolds, closed or not, 
as long as they are orthogonal to $\xiv$, {\it even} if $\xiv$ is not 
hypersurface-orthogonal. It follows that all such 
submanifolds are absolutely non-trapped or extremal (or a combination). 
As a consequence,
in strictly  stationary spacetimes all possible (marginally, nearly) trapped 
surfaces---necessarily non-closed due to theorem \ref{fund}--- must be 
non-orthogonal to $\xiv$. This case can certainly happen---e.g. the 
example of flat spacetime mentioned above---.

Proposition \ref{prop1} and the example just before Theorem \ref{sec}
show that closed extremal submanifolds (i.e. with vanishing mean curvature vector) can exist in static spacetimes when they are contained in a hypersurface
orthogonal to the static Killing $\vec{\xi}$. A natural question is whether 
other types of closed extremal submanifolds can exist. Our last proposition shows that
this is not the case, i.e. that {\it all} closed extremal submanifolds in a  strictly static 
spacetime must lie within a hypersurface of constant static time.

\begin{prop}
Let $\vec{\xi}$ be a strictly static Killing vector on $\espaitemps$ and $S$
a closed spacelike submanifold with vanishing mean curvature vector $\vec{H}=0$. Then,
$S$ is contained in a hypersurface orthogonal to $\vec{\xi}$.
\end{prop}
\proof 
Let $t$ be the static time function of $\espaitemps$, i.e. $\bm{\xi} = - V^2 dt$, where $V^2 =
- ( \vec{\xi} \cdot \vec{\xi} )$. Let also $\tilde{t} \equiv t |_S$, 
$\tilde{V} \equiv V |_S$. Since $\vec{\xi}$ is a
Killing vector it follows, as in the proof of Lemma \ref{res1}, that 
$(\xiv\cdot\vec{H})=-\di \vec{\bar{\xi}}$. 
Since $\bar{\xi}^A= - \tilde{V}^2 D^A \tilde{t}$, where $D^A$
is the covariant derivative of $\gamma_{AB}$, we
immediately derive that
$$
( \vec{H} \cdot \vec{\xi} \, ) =  
\gamma^{AB} D_A \left ( \tilde{V}^2 D_B \tilde{t} \right ).
$$
If $(\vec{H} \cdot \vec{\xi} \, )=0$ (in particular if $S$ is extremal)
we have $\gamma^{AB} D_A ( \tilde{V}^2 D_B \tilde{t} )=0$,
which can be viewed as an elliptic PDE for $\tilde{t}$ on a closed submanifold 
$S$. Uniqueness of solutions for this PDE except for an additive constant 
follows from standard results (see e.g. \cite{GT}).
In our case this can be proven directly
by integrating $D_A ( \tilde{V}^2 \tilde {t} \, D^A \tilde{t} )$
on $S$ and using Gauss' theorem. Consequently
it follows that $\tilde{t} = c$, constant on $S$, which proves that $S$ 
must be contained in a constant
static time slice of the spacetime.  \fin

Many other results may be obtained from formulas (\ref{dereta}) and 
(\ref{vareta}), which can be helpful in other studies too. An 
obvious open question from all the above, that we would like to 
remark, is: in strictly static 
spacetimes, are there any closed 
submanifolds with a causal $\vec H$ everywhere? This question is non-trivial
even in Minkowski spacetime. A positive answer would imply, among other
things, 
that there are closed spacelike surfaces with positive Hawking
mass in flat spacetime. This could have interesting implications
for the use of inverse mean
curvature flows in spacetime as a tool for proving 
the Penrose inequality,
specially if there are closed surfaces
with $\vec{H}$ null everywhere. It would be interesting to settle this issue.
It might be also worth exploring the possible applications of the 
concept of trapped submanifolds of {\em any} codimension to the 
problems in which the codimension two case has been 
important, such as those enumerated at the begining of this letter.

\section*{Acknowledgements} We thank the anonymous referee for
useful comments.
JMMS acknowledges financial support under
grants BFM2000-0018 of the Spanish CICyT and 
no. 9/UPV 00172.310-14456/2002 of the University of the Basque Country.
MM wishes to thank the Junta de Castilla y Le\'on for 
financial support under grant SA002/03


\section*{References}
\begin{thebibliography}{99}


\bibitem{P2} R. Penrose, Gravitational collapse and space-time singularities,
\Journal{\PRL}{14}{57}{1965}.

\bibitem{HP} S.W. Hawking, R. Penrose, The singularities of 
gravitational collapse and cosmology, {\it Proc. Roy. Soc. 
London} A{\bf 314} 529 (1970).

\bibitem{HE} S.W. Hawking, G.F.R. Ellis, {\it The large scale
structure of space-time\/}, (Cambridge Univ. Press, Cambridge, 1973).

\bibitem{S1} J.M.M. Senovilla, Singularity theorems and their 
consequences, \Journal{\GRG}{30}{701}{1998}.

\bibitem{GC} 
P.N. Demni, A.I. Janis, The characteristic development 
of trapped surfaces, \Journal{\JMP}{14}{793}{1973};
U. Alfes, H. M\"uller zum Hagen, Spherically Symmetric Event 
Horizons and Trapped Surfaces Developing from Innocuous Data, 
\Journal{\CQG}{11}{2705}{1994};
R. Beig, N. \'{O} Murchadha, Trapped Surfaces in Vacuum
Spacetimes, \Journal{\CQG}{11}{419}{1994}; R. Beig, N. \'{O} Murchadha,
Vacuum Spacetimes with
Future Trapped Surfaces, \Journal{\CQG}{13}{739}{1996};
P. Bizon, E. Malec, N. \'{O} Murchadha, Trapped surfaces due
to concentration of
matter in spherically symmetric geometries, \Journal{\CQG}{7}{961}{1989}.

\bibitem{CC}
R. Penrose,  Gravitational collapse: the role of general
relativity, \Journal{\NC}{1}{252}{1969};
R. Wald, Gravitational collapse and cosmic censorship, gr-qc/9710068;
D. Christodoulou, The instability of naked singularities in the gravitational 
collapse of a scalar field, \Journal{\AM}{149}{183}{1999};
A. Krolak, Towards the proof of the cosmic censorship hypothesis,
\Journal{\CQG}{3}{267}{1986};
P.T. Chru\'{s}ciel, J. Isenberg, V. Moncrief, Strong cosmic censorship in
polarized Gowdy spacetimes, \Journal{\CQG}{7}{1671}{1990};
M. Dafermos, The interior of charged black holes and the problem 
of uniqueness in general relativity, gr-qc/0307013.

\bibitem{PI}
R. Penrose, Naked singularities, \Journal{\ANYAS}{224}{125}{1973};
G. Gibbons, Collapsing shells and the Isoperimetric inequality, \Journal{\CQG}{14}{2905}{1997};
G. Huisken, T. Ilmanen, The inverse mean curvature flow and the Riemannian 
Penrose inequality, \Journal{\JDG}{59}{353}{2001};
H. Bray, Proof of the Riemannian Penrose inequality using the Positive Mass
Theorem, \Journal{\JDG}{59}{177}{2001};
R. Geroch, Energy Extraction, \Journal{\ANYAS}{224}{108}{1973};
P.S. Jang, R. Wald, The positive energy conjecture and the cosmic censorship,
\Journal{\JMP}{18}{41}{1977};
M. Herzlich, A Penrose-like inequality for the mass of Riemannian
asymptotically flat manifolds, \Journal{\CMP}{188}{121}{1997};
E. Malec, M. Mars, W. Simon, On the Penrose inequality for general horizons,
\Journal{\PRL}{88}{121102}{2002};
M. Ludvigsen, J.A.G. Vickers, An inequality relating total mass and the area of a trapped
surface in general relativity,
 \Journal{\JPA}{16}{3349}{1983};
G. Bergqvist, On the Penrose inequality and the role of
auxiliary spinor fields,
\Journal{\CQG}{14}{2577}{1997}; 
E. Malec, N. \'{O} Murchadha, Trapped surfaces and the Penrose inequality in spherically
symmetric geometries, \Journal{\PRD}{49}{6931}{1994};
S.A. Hayward, Quasi-localization of Bondi-Sachs energy-loss
\Journal{\CQG}{11}{3037}{1994}.


\bibitem{Num}
L. Lehner, Numerical Relativity: a review, \Journal{\CQG}{18}{R25}{2001};
B.K. Berger, Numerical approaches to spacetime singularities, {\it 
Living Rev. Relativity}, lrr-2002-1 (2002);
G.B. Cook, Initial data for numerical relativity, {\it 
Living Rev. Relativity}, lrr-2000-5 (2000).

\bibitem{Kr} M. Kriele, {\em Spacetime}, (Springer, Berlin, 1999).

\bibitem{S2} J.M.M. Senovilla, Trapped surfaces, horizons and exact 
solutions in higher dimensions, \Journal{\CQG}{19}{L113}{2002}.

\bibitem{Jost} J. Jost, {\it Riemannian Geometry and geometric 
analysis}, 3rd ed. (Springer, Berlin, 2001).

\bibitem{O} B. O'Neill, {\em Semi-Riemannian Geometry: 
With Applications to Relativity} (Academic Press, 1983).

\bibitem{C1} R. Bartnik, P.T. Chru\'{s}ciel, N. \'{O} Murchadha, 
On maximal surfaces in asymptotically flat space-times,
\Journal{\CMP}{130}{95}{1990}.

\bibitem{C2} P.T.Chru\'sciel, R.Wald, Maximal hypersurfaces in asymptotically
stationary space-times, \Journal{\CMP}{163}{561}{1994}.

\bibitem{N} R.P.A.C. Newman, Black holes without singularities, 
\Journal{\GRG}{21}{981}{1989}.

\bibitem{Exact} H. Stephani, D. Kramer, M.A.H. MacCallum,
C. Hoenselaers, E. Herlt,  
{\it Exact Solutions to Einstein's Field Equations Second Edition}
(Cambridge University Press, Cambridge, 2003).

\bibitem{CHS} B. Coll B, S.R. Hildebrandt, J.M.M. Senovilla,
Kerr-Schild symmetries, \Journal{\GRG}{33}{649}{2001}.

\bibitem{GS} A. Garc\'{\i}a-Parrado, J.M.M. Senovilla, 
Causal symmetries, \Journal{\CQG}{20}{L139}{2003}; 
General study and basic properties of causal symmetries, gr-qc/0308091.

\bibitem{GT} D. Gilbarg, N.S. Trudinger, {\it Elliptic Partial
Differential Equations of Second Order} (Springer-Verlag, 1983).

\end{thebibliography}
\end{document}